\documentclass[showpacs,prl,aps,superscriptaddress,twocolumn,floatfix]{revtex4}
\usepackage{graphicx,float}

\newfam\ensmathfam                        

\usepackage{epsfig}
\usepackage[latin1]{inputenc}
\usepackage{latexsym}

\def\bea{\begin{eqnarray}}
\def\eea{\end{eqnarray}}
\def\be{\begin{equation}}
\def\ee{\end{equation}}

\begin{document}

\title{Two-component repulsive Fermi gases with population imbalance in elongated harmonic traps}
\author{M. Colom\'e-Tatch\'e}$^1$
\affiliation{
\mbox{Laboratoire de Physique Th\'eorique et Mod\`eles Statistiques, Universit\'e Paris Sud, 91405 Orsay Cedex,France
}}
\date{published 18 September 2008}

\begin{abstract}  

We study the two-component repulsive Fermi gas with imbalanced populations in one dimension. Starting from the Bethe Ansatz solution we calculate analytically the phase diagram for the homogeneous system. We show that three phases appear: the balanced phase, the fully polarised phase and the partially polarised phase. By means of the local density approximation and the equation of state for the homogeneous system we calculate the density profile for the harmonically confined case. We show that a two-shell structure appears: at the center of the cloud we find the partially polarised phase and at the edges the fully polarised one. The radii of the inner and outer shells are calculated for different values of the polarisation and the coupling strength. We calculate the dependence of the magnetisation on the polarisation for different values of the coupling strength and we show that the susceptibility is always finite.
\end{abstract}  
\maketitle
\section{I. INTRODUCTION}
The experimental achievement of a quantum degenerate ultracold Fermi gas \cite{DFG,multi-compo,BEC-BCS crossover} makes it possible to explore experimentally many new models, allowing us to study the rich physics of fermions \cite{review}. At the same time, recent experiments have shown the possibility to study ultra-cold atomic gases confined in very elongated traps \cite{two-d systems,one-d and two-d systems and optical lattice Gorlitz,one-dimensional,one-dimensional and optical lattice}, stimulating theoretical interest in the physics of low-dimensional systems. The one-dimensional regime is reached through tightly confining the radial motion of atoms to zero point oscillations \cite{Olshanii}; this can be done experimentally by loading the atoms in a sufficiently strong optical lattice \cite{one-dimensional and optical lattice} or by means of magnetic wires in an atom chip \cite{magnetic wires}.

Present facilities also allow to create spinor gases by confining in the same trapping potential different atoms or several different internal states of the same atom, allowing us to study the physics of multicomponent atomic systems \cite{multi-compo,BEC-BCS crossover}.\\
In addition, the interatomic interactions can be controlled via a Feshbach resonance \cite{Fano-Feshbach resonance}, making it possible to explore different interaction regimes.

Combining the one-dimensional geometry with the tunable interactions and the possibility of mixing different species of atoms allows us to explore different features such as the Tonks-Girardeau gas \cite{Girardeau,experimental tg gas}, the formation of confinement induced molecules \cite{Olshanii,confinement induced molecules theory,confinement induced molecules Experiments}, the physics of the BEC-BCS crossover \cite{BEC-BCS crossover,J.N.Fuchs}, or the dynamics of one-dimensional systems \cite{dynamics of one-dimensional systems experiments}. Currently, uniform one-dimensional systems are relatively well understood and described. However, the presence of the shallow trapping potential in the long direction of the ultra-cold atomic cloud introduces a source of inhomogeneity in the system, bringing to light new physics which is particularly important in the case of multicomponent atomic systems.

Polarised Fermi gases have been the focus of interest in the recent years \cite{Orso-Drummond-Batchelor,J.N.Fuchs}. In this article we present a study of the exact ground state at zero temperature of a polarised one-dimensional Fermi gas with repulsive inter-component interactions. We show that for the harmonically confined case the atoms are distributed in a two-shell structure: a partially polarised phase is localised in the inner shell and a fully polarised phase sits at the edges of the trap. The radii of the inner and outer shell are calculated as a function of the polarisation. We calculate the dependence of the magnetisation on the polarisation, showing that the system always has a finite susceptibility. We begin by reviewing the Bethe Ansatz solution for the Hamiltonian of the mixture of a two-component Fermi gas with a delta-function repulsive interaction; then we use this solution to calculate the phase diagram of the system in the homogeneous case. Adding a harmonic confinement and using local density approximation, we calculate the properties of the system in the trapped configuration.
\section{II. MODEL} 
Let us consider an atomic cloud confined in an anisotropic harmonic potential 
\be V_{\rm{ext}}(\vec r)=\frac{1}{2}m\omega^2_{\bot}r^2+\frac{1}{2}m\omega^2_{z}z^2, \ee
where $m$ is the mass of an atom and $r=\sqrt{x^2+y^2}$. 
When the ratio between the two trapping frequencies satisfies $\omega_{z}/\omega_{\bot}\ll1$, only the lowest transverse mode is populated; then the low-energy scattering properties of such a system can be modeled by an effective contact interaction in one dimension $U(z)=g_{1D}\delta(z)$ with parameter $g_{1D}=2\hbar^2/ma_{1D}=2\hbar^2a_{3D}/[ma_{\bot}^2\left(1-Ca_{3D}/a_{\bot}\right)]$ \cite{Olshanii},
where $a_{\bot}=\sqrt{\hbar/m\omega_{\bot}}$ is the transverse oscillation length, $a_{1D}$ is the effective 1D scattering length, $C=1.0326...$ is a numerical parameter and $a_{3D}$ is the three-dimensional scattering length. By using magnetic-field-induced Feshbach resonances, the interaction can be tuned from a repulsive effective interaction ($g_{1D}>0$) to an attractive one ($g_{1D}<0$).

We consider a one-dimensional two-component system of spin-$1/2$ fermions interacting via a delta-function potential. The Hamiltonian of the system is
{\small\be
\label{eq:hamiltonian}
H=\frac{\hbar^2}{2m}\left(-\sum_{i=1}^{N}\frac{\partial^2}{\partial z_{i}^2}+2c\sum_{i<j}\delta(z_{i}-z_{j})\right)+V_{\rm{ext}}(z),
\ee}where $c$ is related to the 1D effective interaction coefficient by $c=mg_{1D}/\hbar^2=2/a_{1D}$ and $V_{\rm{ext}}$ is the confining potential. Due to the Pauli exclusion principle, the delta-function interaction only occurs between particles with opposite spin. The total number of particles is $N=N_{\uparrow}+N_{\downarrow}$; without loss of generality we can assume $N_{\downarrow}\leq N/2$. We will call the particles with up (down) spin the majority (minority) component. In one dimension the interactions are measured through a dimensionless coupling constant $\gamma=c/n$, which can be interpreted as the ratio between the mean field interacting energy and the characteristic kinetic energy. $\gamma \ll 1$ corresponds to the weakly interacting regime while $\gamma \gg 1$ stands for the strong interacting one. Hence, in one dimension, the gas is more strongly interacting the lower the density.

In the homogeneous case, without the harmonic confinement, this system was exactly solved by Yang \cite{C.N.Yang} for the case of repulsive interaction using the Bethe Ansatz: an eigenstate of the Hamiltonian Eq.~(\ref{eq:hamiltonian}) is characterised by a set of $N$ density quantum numbers $I_j$ ($j=1, \ldots,N $), and $N_{\downarrow}$ spin quantum numbers $J_{\alpha}$ ($\alpha=1,\ldots,N_{\downarrow}$); which define a set of $N$ quasi-momenta $p_j$ and $N_{\downarrow}$ spin rapidities $\lambda_{\alpha}$. In the thermodynamic limit, they satisfy the set of Bethe Ansatz coupled equations:
{\small\bea
\label{eq:thermodynamic_bethe_ansatz}
\rho(p)&=&\frac{1}{2\pi}+\frac{1}{\pi}\int_{-B}^{B}\frac{2}{1+4(p-\lambda)^2}\sigma(\lambda)d\lambda,\nonumber\\
\sigma(\lambda)&=&\frac{1}{\pi}\int_{-Q}^{Q}\frac{2}{1+4(p-\lambda)^2}\rho(p)dp\nonumber\\&-&\frac{1}{\pi}\int_{-B}^{B}\frac{1}{1+4(\lambda-\lambda')^2}\sigma(\lambda')d\lambda',
\eea}
where all the parameters have been rescaled by $c$: $\lambda=\lambda/c$, $p=p/c$, $B=B/c$, $Q=Q/c$. $\rho(p)$ should be understood as the density of $p$ (density of particles) and $\sigma(\lambda)$ as the density of $\lambda$ (density of down-spin particles). The cut-offs $B$ and $Q$ are determined by fixing the density of down-spin particles ($n_{\downarrow}$) and the total density of particles ($n=n_{\downarrow}+n_{\uparrow}$) in the system:
\be
\label{eq:number_particles}
\frac{n}{c}=\int_{-Q}^{+Q}\rho(p)dp,\;\;\;
\frac{n_{\downarrow}}{c}=\int_{-B}^{+B}\sigma(\lambda)d\lambda.
\ee
The weak coupling regime corresponds to $n/c\gg1$ and the strong coupling regime to $n/c\ll1$.
The magnetisation is defined as $s=n_{\uparrow}-n_{\downarrow}$, and the expression for the energy reads
\be 
\label{eq:thermodynamic_energy}
\frac{E(n,s)}{L}=\frac{\hbar c^3}{2m}\int_{-Q}^Q p^2 \rho(p)dp.
\ee 
\section{III. PHASE DIAGRAM} 
We will now calculate the phase diagram for the system. In order to do so, we fix the total density of particles $n$ and change the balance between the numbers of up and down spins, that is to say the magnetisation $s$. When $s=0$, the system is at balance, since the number of up spins equals the number of down spins, $n_{\uparrow}=n_{\downarrow}$. Since the excitation spectrum is gapless this phase only exists in the situation with zero magnetic field. For $s=n$ the system is fully polarised and $n_{\downarrow}=0$. For any intermediate value of the magnetisation $0<s<n$ the system is an imbalanced mixture of up and down spins, i.e. partially polarized. There is no evidence of a finite momenta instability, such as in the FFLO state \cite{FFLO}, in the partially polarised phase: a calculation of the critical exponents for the pair correlation function shows that this correlation is never the leading one for the case with repulsion \cite{Frahm_Korepin}, contrary to what happens in the case of attraction \cite{KunYang}.

The magnetic field and the chemical potential,
\be
\label{eq:definition_h_mu}
h=\frac{\partial E(n,s)/L}{\partial s},\;\;\;\mu=\frac{\partial E(n,s)/L}{\partial n},
\ee
can be calculated from Eq.~(\ref{eq:thermodynamic_bethe_ansatz}),~(\ref{eq:number_particles}) and ~(\ref{eq:thermodynamic_energy}): the energy depends directly on the cut-offs $Q$ and $B$, so one has to take the derivative of ~(\ref{eq:thermodynamic_energy}) with respect to $Q$ and $B$ and then use ~(\ref{eq:number_particles}) to find the derivatives of $Q$ and $B$ with respect to $n$ and $s$. The magnetic field and chemical potential can also be related to the chemical potential of the two species of fermions as $h=\frac{\mu_{\uparrow}-\mu_{\downarrow}}{2}$ and $\mu=\frac{\mu_{\uparrow}+\mu_{\downarrow}}{2}$.

Setting $s=n$, or equivalently $n_{\downarrow}=0$ in Eq.~(\ref{eq:definition_h_mu}), one can calculate the boundary between the mixed imbalanced phase, where $0<s<n$, and the fully polarised phase corresponding to $s=n$. The saturation magnetic field and the corresponding chemical potential for which the system becomes fully polarised are
\bea
\label{eq:critical_line}
h_s&=&2\epsilon_B\left(\frac{Q_0^2}{\pi}\arctan(2Q_0)-\frac{Q_0}{2\pi}+\frac{\arctan(2Q_0)}{4\pi}\right),\nonumber\\
\mu_s&=&-h+2\epsilon_B Q_0^2,
\eea
where $\epsilon_B=\hbar^2 c^2/4m$ is the binding energy and $Q_0=n\pi/c$ is the Fermi momentum.

\begin{figure}[t]
\centerline{
	\mbox{\includegraphics[width=3.5in]{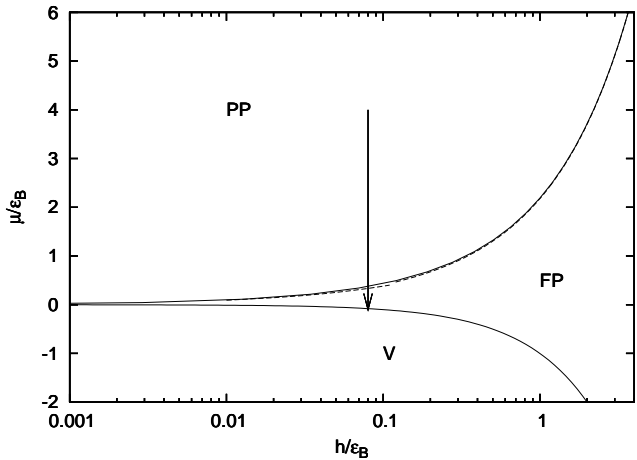}}
  }
  \caption{Phase diagram in the plane $\mu/\epsilon_B$, $h/\epsilon_B$. The partially polarised phase (PP), fully polarised phase (FP) and vacuum (V) are displayed. The dashed lines correspond to the asymptotes discussed in the text; the vertical arrow is discussed in the frame of the Local Density Approximation.}
  \label{Fig.1}
\end{figure}

In the strong coupling limit ($Q_0\rightarrow 0 \Leftrightarrow h\ll \epsilon_B$) the chemical potential tends to zero as 
\be
\label{eq:critical_line_zero} \frac{\mu}{\epsilon_B}=-\frac{h}{\epsilon_B}+2\left(\frac{3\pi}{8}\right)^{2/3}\left(\frac{h}{\epsilon_B}\right)^{2/3};\ee
while for the weak coupling limit ($Q_0\rightarrow \infty \Leftrightarrow h\gg \epsilon_B$) the chemical potential diverges according to
\be \label{eq:critical_line_diverge} \frac{\mu}{\epsilon_B}=\frac{h}{\epsilon_B}+\frac{4}{\pi}\sqrt{\frac{h}{\epsilon_B}}+\left(\frac{4}{\pi^2}-\frac{1}{2}\right),\ee 
which recovers the result obtained using a mean-field approach. Since the excitation spectrum is gapless, the saturation line crosses the point $h=0,\mu=0$. The vacuum line, for which $n=0$, corresponds to $\mu_v=-h$. The phase diagram is shown in Fig.~\ref{Fig.1}.

We will now calculate the magnetic susceptibility close to saturation. Using the equality 
{\small\be
\int_{-B}^{B}f(\lambda)\sigma(\lambda)d\lambda=f(\lambda^*)\int_{-B}^{B}\sigma(\lambda)d\lambda,\; -B<\lambda^*<B,
\ee}we can rewrite $Q$ (\ref{eq:number_particles}) and the energy (\ref{eq:thermodynamic_energy}) as:
{\small{\setlength\arraycolsep{2pt}\bea
\label{eq:Q_small_n-s}
Q&=&Q_0-\frac{n-s}{c}\arctan(2Q)\\
\label{eq:E_small_n-s}
\frac{E(n,s)}{L}&=&\frac{\hbar^2c^3}{2m}\left(\frac{Q^3}{3\pi}+\frac{n-s}{c}\frac{1}{2\pi}\int_{-Q}^{Q}\frac{2p^2}{1+4p^2}dp\right).
\eea}}
After solving $Q$ in (\ref{eq:Q_small_n-s}) up to second order in the small parameter $(n-s)$, we can calculate the energy near saturation using (\ref{eq:E_small_n-s}). Taking the first derivative of the energy with respect to the magnetisation and the number of particles at $n=s$ we recover the result for the magnetic field and the chemical potential at saturation (\ref{eq:critical_line}). Taking the second derivative of the energy with respect to the magnetisation we obtain the value of the magnetic susceptibility at saturation:
\be 
\label{eq:susceptibility_saturation}
\chi^{-1}_s={\textstyle \left.\frac{\partial^2 E(n,s)/L}{\partial s^2}\right|_{s=n}}=\frac{2\epsilon_B}{c} \left(\frac{2Q_0}{\pi} \arctan^2(2Q_0)\right),
\ee
which only diverges in the limit of the empty band ($Q_0=0$) and in the limit of infinite interaction, and remains finite otherwise. We want to stress that the calculations have been performed for the case with fixed number of particles, for which the van Hove singularity is not present, therefore we do not expect a divergency of the magnetic susceptibility at saturation.
\section{IV. TRAPPED DENSITY PROFILES} 
\begin{figure}[t]
\centerline{
	\mbox{\includegraphics[width=2.8in]{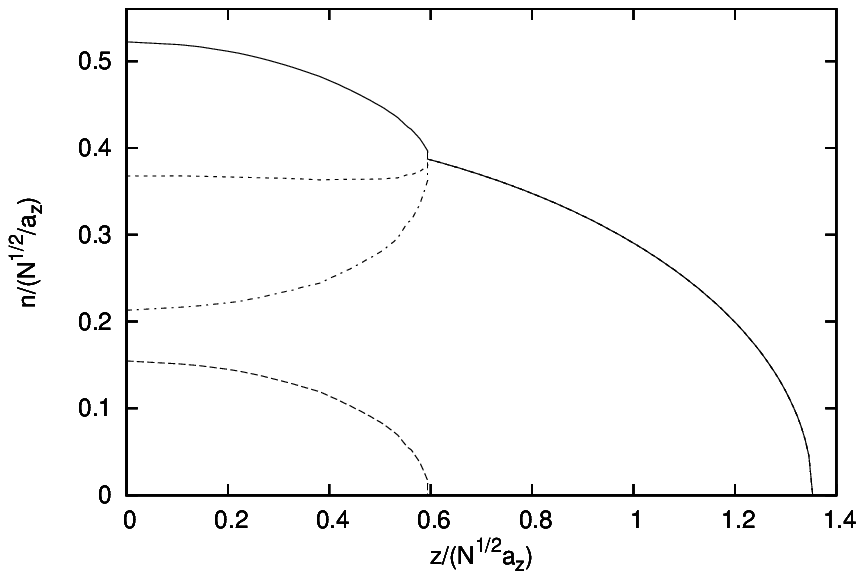}}
}\centerline{
	\mbox{\includegraphics[width=2.8in]{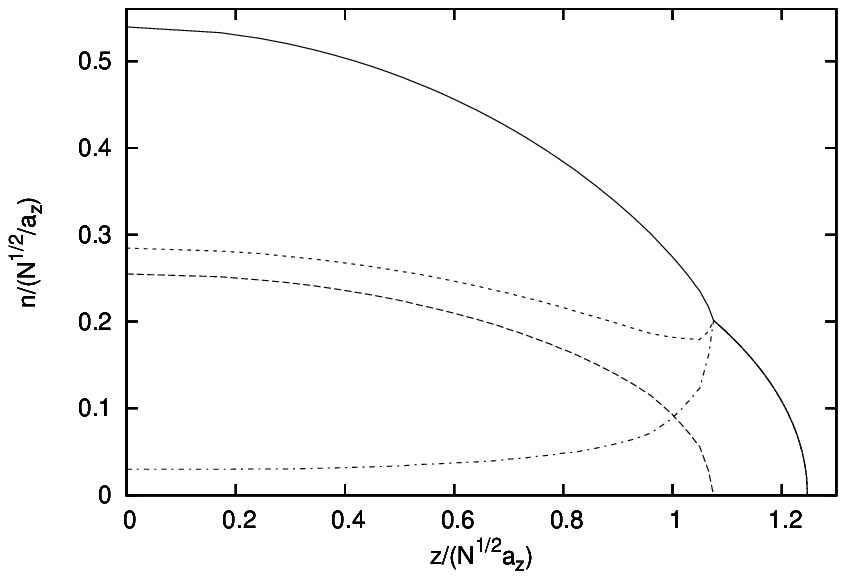}}
 }
  \caption{Density profile for polarisation $P=0.7$ (top) and $P=0.15$ (bottom) for $N(a_{1D}^2/a_{z}^2)=1$. The solid line is the total density
$n(z)/(\sqrt{N}/a_{z})$, the dotted line is the density of up spins
$n_{\uparrow}(z)/(\sqrt{N}/a_{z})$, the dashed line is the density of down spins
$n_{\downarrow}(z)/(\sqrt{N}/a_{z})$ and the dashed-dotted line is the magnetisation 
$m(z)/(\sqrt{N}/a_{z})$. We can see the two shells: the inner one contains an imbalanced mixture of up and down spins, while the external shell is fully polarised.}
\label{Fig.2}
\end{figure}

Let us now add a harmonic confinement along the axial direction $V_{\rm{ext}}(z)=m\omega_z^2z^2/2$. If the trap is sufficiently shallow then the size of the cloud can be considered much bigger than $a_z$ and one can treat the system as being locally uniform. This allows us to use the Local Density Approximation (LDA) in order to write the local equilibrium conditions
\bea
\label{eq:L_D_A}
\mu(n_{\uparrow}(z),n_{\downarrow}(z))&=&\mu^o-\frac{1}{2}m\omega_z^2z^2,\nonumber\\
h(n_{\uparrow}(z),n_{\downarrow}(z))&=&h^o,
\eea
where 
\bea &\mu(n_{\uparrow}(z),n_{\downarrow}(z))=\frac{\mu_{\uparrow}(n_{\uparrow}(z),n_{\downarrow}(z))+\mu_{\downarrow}(n_{\uparrow}(z),n_{\downarrow}(z))}{2},&\nonumber\\ &h(n_{\uparrow}(z),n_{\downarrow}(z))=\frac{\mu_{\uparrow}(n_{\uparrow}(z),n_{\downarrow}(z))-\mu_{\downarrow}(n_{\uparrow}(z),n_{\downarrow}(z))}{2}&
\eea
are the chemical potentials calculated for the homogeneous system, and
$\mu^o=(\mu_{\uparrow}^o+\mu_{\downarrow}^o)/2$ and $h^o=(\mu_{\uparrow}^o-\mu_{\downarrow}^o)/2$ are constants calculated imposing the normalisation conditions for the number of particles in each component, i.e. for the total number of particles $N$ and the total magnetisation $S$
\bea
\label{eq:normalisation_LDA}
N=\int_{-R}^{R} n(z) dz,\;S=\int_{-R}^{R} s(z) dz.
\eea
Note that in our situation the two species are confined in the same potential (since we can consider them as being two different hyperfine states of the same atom) and so the local magnetic field $h(z)$ is kept constant along the trap, while the local chemical potential $\mu(z)$ decreases as we approach the edges of the cloud (see arrow in Fig.~\ref{Fig.1}). In order to calculate the density profile, we solve numerically Eq.~(\ref{eq:L_D_A}) fixing the total number of particles and the magnetisation:
{\small\bea
N\frac{a_{1D}^2}{a_z^2}=2\sqrt{2}\int_{-\tilde{R}}^{\tilde{R}}\tilde{n}(\tilde{z})d\tilde{z},\nonumber\\
S\frac{a_{1D}^2}{a_z^2}=2\sqrt{2}\int_{-\tilde{R}}^{\tilde{R}}\tilde{s}(\tilde{z})d\tilde{z},
\eea}i.e. fixing the values for $\mu^o$ and $h^o$. Here $\tilde{z}=z a_{1D}/\sqrt{2}a_z^2$, $\tilde{n}=n/c$ and $\tilde{s}=s/c$. The density profiles for different values of the polarisation $P=(N_{\uparrow}-N_{\downarrow})/N$ are shown in Fig.~\ref{Fig.2}.

We can understand the density profile as follows: when moving along the trap, the chemical potential difference $h(z)$ remains constant while the chemical potential $\mu(z)$ decreases parabolically and so does the one-dimensional local density. The interaction strength $\gamma(z)=c/n(z)$ increases while we move away from the center of the cloud, since it is inversely proportional to the one-dimensional local density. The system is gapless, therefore for any applied magnetic field $h$ it is in the mixed phase, that is to say in a mixture of up and down spins, $s\neq0$. That explains why moving from the center of the trap towards the edges of the cloud we find two different phases: at the center we have an imbalanced mixed phase for any magnetic field $h$, while at the edges, where the two species become more repulsive, we find a fully polarised phase.

In three dimensions the density profiles at weak interaction strength present a very similar qualitative structure \cite{3DProfiles}. However, when increasing the interaction strength in three dimensions a symmetry breaking occurs, driven by the competition between the repulsive interaction energy and the kinetic energy, and a phase separation occurs: the minority component is pushed to the edges of the trap while the majority component accommodates in the center \cite{3DProfiles}. In one dimension we can calculate the limit of infinite inter-species interaction exactly by means of the exact mapping to a fermionic Tonks-Girardeau gas (two-component Fermi gas with hard-core inter-species interaction) for which the density profile is known to be equivalent to that of $N=N_{\uparrow}+N_{\downarrow}$ bosons with Tonks-Girardeau point-like interaction \cite{Girardeau-Minguzzi}. There is though no phase-separation expected in this limit in the one-dimensional configuration, which is consistent with our numerical calculations.
\begin{figure}[t]
  \centerline{
\hspace{0.7cm}
\mbox{\includegraphics[width=3.45in]{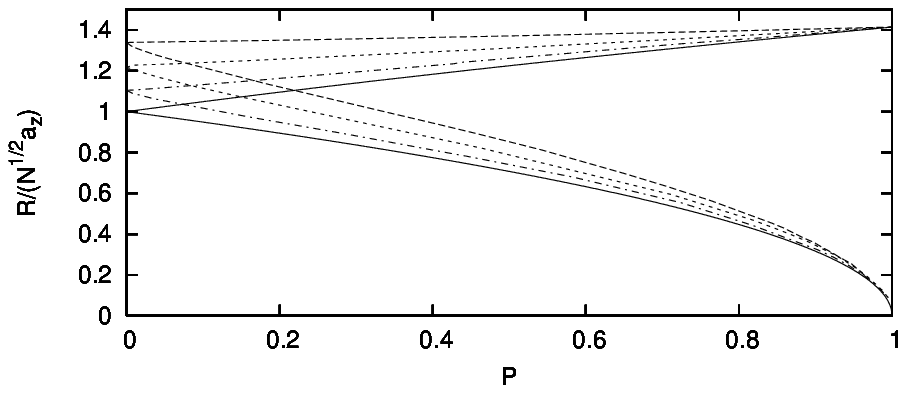}}
 }
	\vspace{-5.2cm}
  \centerline{
	\mbox{\includegraphics[width=3.25in]{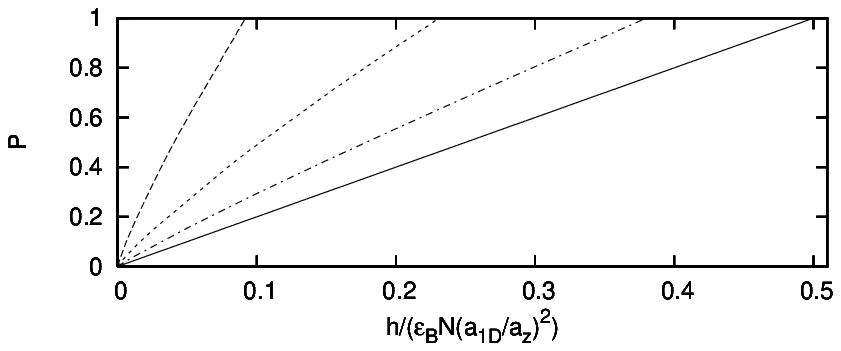}}
  }
  \caption{Top: Radii of the inner (inferior line) and the outer (superior line) shells. Bottom: phase diagram in the plane $P$ vs. $h/\epsilon_B$. For $N(a_{1D}^2/a_{z}^2)=\infty$ (solid line), $N(a_{1D}^2/a_{z}^2)=10$ (dashed-dotted line), $N(a_{1D}^2/a_{z}^2)=1$ (dotted line) and $N(a_{1D}^2/a_{z}^2)=0.1$ (dashed line).
  }
  \label{Fig.3}
\end{figure}

We now calculate the radii of the inner and outer shells in the density profiles. The first is given by $(a_{1D}/a_{z}^2)R_{in}=\sqrt{2\left(\mu^o/\epsilon_B-\mu_s/\epsilon_B\right)}$ and the last by $(a_{1D}/a_{z}^2)R_{out}=\sqrt{2\left(\mu^o/\epsilon_B-\mu_v/\epsilon_B\right)}$. We compute them as a function of the polarisation for different values of the interaction strength (Fig.~\ref{Fig.3}). We see perfect agreement between the numerical results and the limiting analytical result for free fermions, which is displayed in Fig.~\ref{Fig.3} with solid lines. We also see agreement with the limiting case of infinite repulsion: this limit is singular, since the interaction between the different species mimics the Pauli principle and we can no longer differentiate between up and down spin particles. In this case, the radius of the cloud is $\sqrt{2N}a_z$ and is independent of the polarisation \cite{Girardeau-Minguzzi}. We also display in Fig.~\ref{Fig.3} the variation of the polarisation with the magnetic field for a fixed number of particles. The onset of magnetisation occurs at $h=0$, proving the absence of a gap in the system. The slope of the magnetisation at the onset of polarisation is sensitive to the rate at which the gas polarises; the linear behaviour is a sign of a finite susceptibility \cite{Woynarovich}. The slope remains close to a constant in the regime $0<h<h_s$, showing that the polarisability of the system is very weakly affected by the extra imbalanced electrons. Even though there is a discontinuity in the slope of the polarisation versus the magnetic field at the saturation point, we do not expect a divergency of the magnetic susceptibility at saturation (\ref{eq:susceptibility_saturation}).
\section{V. CONCLUSIONS}
In conclusion, we have given an exact solution for the phase diagram of the ground state of a one-dimensional two-component Fermi gas with inter-component repulsive interactions, for any value of the coupling strength. Three phases are localised: a balanced phase that only exists at zero magnetic field, a phase with an imbalanced mixture of the two components, and a fully polarised phase containing only the majority component. We show that inside a harmonic trap the cloud presents a double-shell structure, with an imbalanced mixture of the two components in the center and a fully polarised edge. The radii of the inner and outer shells are calculated for different values of the polarisation and the coupling strength. We finally calculate the dependence of the magnetisation on the polarisation for a fixed number of particles, and we show that the susceptibility never diverges.

This model is experimentally accessible. One can cool ${}^{40}\textrm{K}$ atoms to quantum degeneracy, then using radio-frequency pulses prepare a spin mixture of the atoms in the different spin states $\left|F=9/2,m_F=-9/2\right\rangle$ and $\left|F=9/2,m_F=-7/2\right\rangle$ with different concentrations. Adiabatically superposing a 2D optical lattice leads to arrays of nearly identically parallel 1D traps.
\section{ACKNOWLEDGEMENTS} 
The author would like to thank G.V. Shlyapnikov, P. Pedri, G. Orso, T. Vekua, D. Smith, D.M. Gangardt, A. Minguzzi and X. W. Guan for fruitful discussions. The author also wants to acknowledge B. Marcelis and D.S. Petrov for useful suggestions on numerical methods. The work was supported by the IFRAF Institute and by ANR (grants 05-BLAN-0205 and 06-NANO-014-01). LPTMS is a mixed research unit No. 8626 of CNRS and Universit\'e Paris Sud.

\end{document}